# Social Engagement of Children with Autism during Interaction with a Robot

**Adriana Tapus**
Cognitive Robotics Lab
ENSTA-ParisTech, France
adriana.tapus@ensta-paristech.fr

**Andreea Peca, Amir Aly, Cristina Pop, Lavinia Jisa, Sebastian Pintea, Alina Rusu, and Daniel David**

1. Introduction

Imitation plays an important role in development, being one of the precursors of social cognition. Even though some children with autism imitate spontaneously and other children with autism can learn to imitate, the dynamics of imitation is affected in the large majority of cases. Existing studies from the literature suggest that robots can be used to teach children with autism basic interaction skills like imitation.

Based on these findings, in this study, we investigate if children with autism show more social engagement when interacting with an imitative robot (Fig 1) compared to a human partner in a motor imitation task.

2. Method

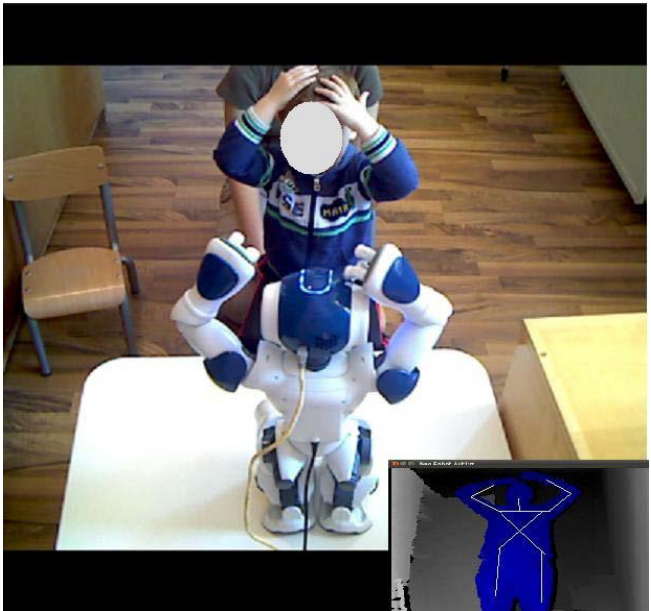

A single-subject ABAC design with replication across 4 participants was employed. The independent variable was the type of interaction agent: robot versus human person. The dependent variables measured were: frequency of free initiations (without prompt), frequency of total initiations (with and without prompt) and imitations, duration of attention to the partner (robot/person), duration of smile, and frequency of gaze shifting between the two partners (robot-therapist/human person-therapist).

Fig 1: Child #2 interacting with the robot

3. Results

The results confirm only partially the experimental hypothesis. The Mann -Whitney test was used to analyze the difference between each pair of conditions. Child #1 and Child

#2 showed more attention and positive affect in the interaction with Nao robot. In the case of the initiations and imitations variable, the robot was superior for Child #1 and Child #3. The robot proved to be a better facilitator of shared attention only for Child #1 (Fig 2). No significant results were found for free initiations. However, these results are not conclusive, because of the small number of participants used in the study, but are promising for the future directions of this work.

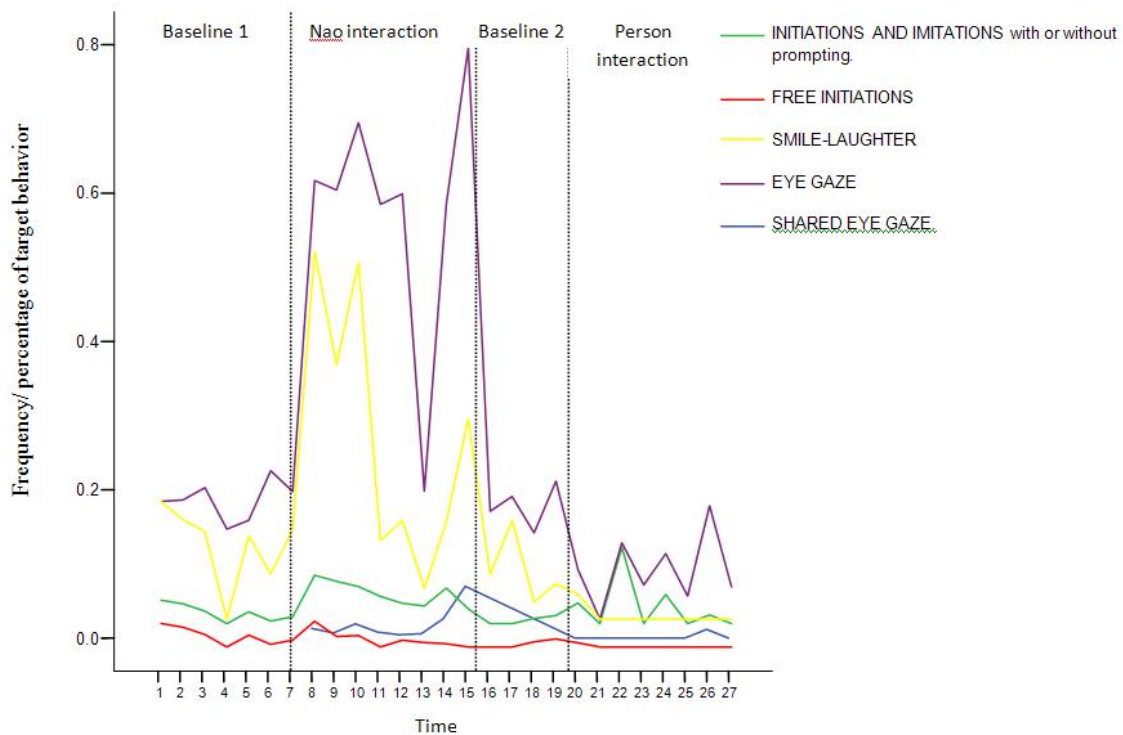

Fig 2: Results Child #1

4. Conclusion

We consider that the major benefit of this study stands in the steps taken towards adapting standard single- subject paradigms into human-robot interaction. This kind of study represents a small step in advancing the engineering field towards the goal of actually providing a tool for clinicians and therapists.